\RequirePackage{fix-cm}
\documentclass{svjour3}                     % onecolumn (standard format)
\smartqed  % flush right qed marks, e.g. at end of proof
\usepackage{graphicx}
\usepackage{fancyhdr}
\usepackage{arydshln}
\usepackage{rotating}
\usepackage{float,lscape}
\usepackage{color}
\usepackage[final]{changes}
\begin{document}

\title{Influence measures in subnetworks using vertex centrality %\thanks{Grants or other notes
%about the article that should go on the front page should be
%placed here. General acknowledgments should be placed at the end of the article.}
}
%\subtitle{Do you have a subtitle?\\ If so, write it here}

%\titlerunning{Short form of title}        % if too long for running head

\author{Roy Cerqueti \and Gian Paolo Clemente         \and
        Rosanna Grassi
}

%\authorrunning{Short form of author list} % if too long for running head

\institute{Gian Paolo Clemente \at
              Department of Mathematics for Economics, Financial and Actuarial Sciences \\
              Universit\'a Cattolica del Sacro Cuore, Milano\\
              \email{gianpaolo.clemente@unicatt.it}         %  \\
%             \emph{Present address:} of F. Author  %  if needed
           \and
           Roy Cerqueti \at
           Department of Economics and Law\\
           University of Macerata,
           Via Crescimbeni 20, 62100, Macerata, Italy.\\
           Phone: +39 0733 2583246\\
           \email{roy.cerqueti@unimc.it}
           \and
			Rosanna Grassi, corresponding author     \at
            Department of Statistics and Quantitative Methods\\
          University of Milano-Bicocca,
              Via Bicocca degli Arcimboldi, 8, 20126 Milan, Italy\\
              Phone: +39-02-64483136\\
              \email{rosanna.grassi@unimib.it}
}

\date{Received: date / Accepted: date}
% The correct dates will be entered by the editor

\fancyhead[RO,LE]{\small\thepage}
\fancyhead[LO]{\small Influence measures in subnetworks using vertex centrality}% odd page header and number to right top
\fancyhead[RE]{\small R. Cerqueti, G.P. Clemente, R. Grassi}%Even page header and number at left top
\fancyfoot[L,R,C]{}
\maketitle

\begin{abstract}
This work deals with the issue of assessing the influence of a node in the entire network and in the subnetwork to which it belongs as well, adapting the classical idea of vertex centrality. We provide a general definition of relative vertex centrality measure with respect to the classical one, referred to the whole network. Specifically, we give a decomposition of the relative centrality measure by including also the relative influence of the single node with respect to a given subgraph containing it.
The proposed measure of relative centrality is tested in the empirical networks generated by collecting assets of the $S\&P$ 100, focusing on two specific centrality indices: betweenness and eigenvector centrality. The analysis is performed in a time perspective, capturing the assets influence, with respect to the characteristics of the analysed measures, in both the entire network and the specific sectors to which the assets belong. %The empirical study sheds light on the behaviour of some assets over time in the market.

% \PACS{PACS code1 \and PACS code2 \and more}
% \subclass{MSC code1 \and MSC code2 \and more}
%\noindent\textbf{JEL classification code}{: XXXXX.}

\end{abstract}

\section{Introduction}
\label{intro}
Complex networks are experiencing an increasing popularity among scientists, either under a methodological as well as practical perspective. They represent a versatile framework for the description of real-world systems with interconnected components (see e.g. \cite{Newman2010}, \cite{WasFaust}).
In the context of complex networks, a very relevant theme is the assessment of the relevance of the single nodes in the overall structure. In this respect, we mention e.g. \cite{Ferraro2017}, \cite{Cerqueti2018}, \cite{Ma2019} and \cite{Wang2011}, where the identification of the key actors among the agents is a crucial task for exploring the proposed applied problem -- inter organizational innovation, social media and air transportation, respectively.

Widely used instruments for identifying the influence of the single nodes in a complex network are the so-called centrality measures. Such devices compound a set of methodological tools sharing the same target of measuring the relevance of the nodes, with the distinctions due to the declination of the concept of \emph{relevance} (see e.g. \cite{Freeman, Perra2008, Watts2004}).

Centrality measures are usually defined as absolute quantities, hence providing an objective description of the importance of the individual nodes of the network. Essentially, they can be also presented as normalized terms, so that nodes -- also belonging to different networks -- can be compared according to their relevance/centrality measure. This universality property of the definition of centrality measures has the severe drawback of not allowing the contextualization of the nodes relevance in the overall network. To fix ideas, think at a node with a very high degree, namely a hub. If such a node belongs to a network with low average degree, then the network is star-shaped and the hub is the crucial node; if, contrastingly, such a node belongs to a network with high average degree, then the considered hub is an important element, but it is not the only one. 
%More details can be given in the latter situation.
For instance, the hub can be part of a rich club (see e.g. \cite{Cinelli2017, Cinelli2018, Colizza2006, OpsahlNEt}), i.e. of a proper subset of nodes with high degree, or it can be ``one among many'', because even all the other nodes of the network have high degree. All these aspects are not covered by the absolute centrality measures.

Therefore, although traditional centrality measures have been formulated for individual nodes, it is equally interesting to explore the idea of a group centrality.
As pointed out in \cite{Everett1999}, the group centrality allows to ``quantify'' the membership of a node to a group. For instance, this could be useful to efficiently remodulate groups, removing internal redundant ties that poorly contribute to the group importance.

According to the arguments above, this paper adds to the debate on centrality measures by proposing two natural advancements to the related theory. First, it provides a general definition of the relative centrality measure of a node with respect to the classical one of the entire network. Second, it offers a decomposition of the relative centrality measure by including also the relative influence of the single node with respect to a given subgraph containing it, hence leading to a concept of group centrality.

One intuitive approach to define a group centrality is to average the centrality scores in the group, but more suitable and effective group centrality measures have been proposed (\cite{Everett1999}). In this work, we consider, in the same formula, the average centrality of nodes in a group and the group centrality defined in the literature. In particular, the comparison between these two values allows to catch the effect of the external vertices to the centrality measure of the nodes belonging to a specific group.

Our final target is then to quantify the importance of a vertex with respect to a subnetwork. This importance will be measured in terms of centrality. In other words, we aim at catching how much a vertex has a central role with respect to both the whole graph and the subgraphs to which the vertex belongs.

In assessing subgraph centrality, we are in line with \cite{Estrada2005}, where the authors propose a characterization of the nodes on the basis of the loops containing them. Such closed walks can be identified with related connected subgraphs, so that their number proxies the relevance of the single nodes over the subgraphs of the network.

We here adopt a different perspective by dealing with a relative measure rather than an absolute one; in this way, our approach allows the comparison of nodes and subgraphs also in presence of different networks.

The theoretical model is validated through empirical experiments based on the daily returns of the components of the $S\& P$ 100 for the period Jan 1st, 2001 - Dec 31st, 2017. A system of networks is considered on the ground of a time windows analysis. The arcs are weighted through the correlation coefficients between couples of assets in the specific time windows and nodes are the assets. The validation is carried out in the context of two specific relative centrality measures: betweenness and eigenvector centrality. The former one gives information on how nodes are relevant in terms of their role in connecting other nodes of the graph, and it has been introduced by \cite{Freeman1977}; the latter centrality measure -- whose introduction dates back to the end of the nineteenth century, and we refer to \cite{Bonacich1987} -- assigns a high power to the nodes connected to highly relevant nodes. 
We purposely focused on two alternative centrality measures with a different meaning. Both measures overcome the simple degree centrality, that refers exclusively to the node's neighbours.  On one hand, the betweenness score catches how a node is influential in controlling the flow of information along shortest paths in the network. On the other hand, the eigenvector centrality captures influences at long distances. We argue that these differences can seize the hidden role of assets in local communities. Moreover, both measures represent suitable tools for evaluating nodes' role in large networks.
%\added{We purposelly focused on two measures \color{red}with atotally different in the meaning. Both measures go one step further the simple degree centrality, that refers exclusively to the node's neigbours. On one hand, the betweenness score measures how a node is influent in controlling the flow of information along shortest paths in the network. On the other hand, the eigenvector centrality captures influences at long distances. We argue that these differences can bring out the hidden role of assets in local communities. Moreover, both measures are often used for large networks.}
The empirical analysis is carried out in a time perspective, capturing the assets influence, with respect to the characteristics of the analysed measures, in both the entire network and the specific sectors to which they belong. Main results show that such measures are of particular interest in the proposed exercise and offer important insights on the reality of the considered empirical sample.

The rest of the paper is organized as follows. Section \ref{sec:prelnot} outlines the notation used in the paper, with the basic concepts. Section \ref{sec:c} is devoted to the theoretical formalization of the relative centrality of nodes and subgraphs in a very general environment. Section \ref{sec:emp} contains the empirical validation of the theoretical model. Such a section is divided in subsections, with the aim of giving a detailed view of the considered dataset, on the specific relative centrality measures employed for the exercise along with some remarks on limitations and comparisons with other measures. Section \ref{sec:res} presents and discusses critically the results of the empirical experiments. Last section offers some conclusive remarks and traces lines for future research.

\section{Preliminaries and notations}
\label{sec:prelnot}
We now review some theoretical concepts about graphs and networks \footnote{For a detailed treatment we refer, for instance, to \cite{Harary} and \cite{Newman2010}.}. Formally, a network is represented by a graph $G=(V,E)$, that is a set of $n$ nodes (vertices) $V$ and $m$ edges $E$ of unordered pairs of vertices.
Two nodes are adjacent if there is an edge $(i,j)$ connecting them.
$G$ is undirected if $(j,i)\in E$ whenever $(i,j)\in E$.
The complete graph $K_n$ is the graph in which every pair of distinct vertices is linked by an edge. A $i-j$ path is a sequence of distinct adjacent vertices from vertex $i$ to vertex $j$. The distance $d(i,j)$ between $i$ and $j$ is the length of the shortest path joining them when such a path exists, and it is set to $+\infty$ otherwise.
A graph $G$ is connected if there is a path between every couple of vertices. A subgraph $G_s=\left( V_s,E_s\right) $ of $G$ is a graph such that $V_s\subseteq V$ and $E_s\subseteq E$. A particular class of subgraphs is the one of the induced subgraphs. A subgraph $G_s=\left( V_s,E_s\right) $ of $G$ is induced by $V_s$ when $(i,j) \in E$ implies $(i,j) \in E_s$, for each $i,j \in V_s$. A maximal connected subgraph of $G$ is called connected component of $G$. A graph is connected if has exactly one connected component.

In general, the adjacency relationships between vertices of $G$ are described by a nonnegative, real $n$-square matrix $\mathbf{A}$ (the
adjacency matrix). We denote with $\rho$ its spectral radius and $\mathbf{x}$ the associated eigenvector. If $G$ is connected, then $\mathbf{A}$ is irreducible and, by Perron Frobenius Theorem, all elements of the eigenvector associated with the spectral radius are strictly positive. This eigenvector is called Perron (or principal) eigenvector.

\section{Relative centrality of a subgraph}
 \label{sec:c}
In the following, we assume that $G=(V,E)$ is a connected %unweighted
and undirected graph of $n$ nodes. For our purposes, we exclude the case of %unweighted
complete graph $K_n$, as all vertices show in this case the same topological structure\footnote{For many centrality measures proposed in the literature, a closed formula computing the centrality of a vertex in a complete graph $K_n$ is provided. }.

Centrality is generally defined in terms of a function $c: V \rightarrow[0,+\infty)$, that assigns nonnegative real values to nodes of the set $V$ of a graph such that
\begin{equation}
 c(i) \geq c(j)  \iff  i \:  \textnormal{is at least central as} \: j.
\label{orderc}
\end{equation}

Without losing of generality, we assume that centrality measures are normalized, so that $c(i) \in [0,1]$ for each $i \in V$.\\ %In the following, we assume that the centrality measures we deal with are normalized.
%The most used centrality measures are betweenness, closeness, degree and eigenvector centrality. For each of them,
According to this condition, we introduce the relative incidence of the centrality of a node $i$ with respect to the average centrality of the graph $G$ (or, simply, \textit{relative centrality of $i$ with respect to $G$}) as:
\begin{equation}
c(i|G)=\frac{c(i)}{{\bar{c}(G)}}
\label{rel}
\end{equation}
where ${\bar{c}(G)}=\frac{\sum_{j =1}^n c(j)}{n}$ is the average centrality of $G$. \\

This index allows a direct comparison of the relative centralities of the specific nodes when considering also the overall related graphs.

%Assume, for instance, to compare two graphs with a different average betweenness.
By means of a relative index we can assess how much a vertex is ``relevant'' in a network, where the peculiar declination of the concept of relevance depends on the specific centrality measure employed. In a general sense, the centrality of the node is compared with the average centrality of the graph. In this respect, notice that $c(i|G)$ can be lower or higher than $1$, according to its position with respect to the average behaviour of the network.

The total order in (\ref{orderc}) can be reproduced also in the relative case, so that
\begin{equation}
 c(i|G) \geq c(j|G)  \iff  i \:  \textnormal{is at least relatively central as} \: j.
\label{ordercrel}
\end{equation}

It is worth noting that this definition does not alter the centrality ranking. Indeed, comparing $c(i|G)$ through order in (\ref{ordercrel}) leads to the same order of comparing $c(i)$ through order in (\ref{orderc}).

The definition of $c(i|G)$ in (\ref{rel}) allows to investigate also the role/position of a vertex with respect to both the whole network and any subnetwork having it as a node.
Let $G_s=\left(V_{s},E_{s}\right)$, be a subgraph of $G$ (where the cardinality of $V_s$ is $n_s$) which $i$ belongs to. For instance, $G_s$ could be the induced subgraph of $n_s \leq n$ nodes of $G$. Moreover, assume that there exists $j \in V_s$ such that $c(j) >0$.

Then, formula (\ref{rel}) can be rewritten as:
\begin{equation}
c(i|G)=\frac{c(i)}{{\bar{c}(G_s)}}\frac{\bar{c}(G_s)}{{\bar{c}(G)}}=c(i|G_s )r_{G_s}
\label{rel2}
\end{equation}
where ${\bar{c}(G_s)}=\frac{\sum_{j \in V_s}c(i)}{n_s}$ is the average centrality of $G_s$. %Notice that, since ${\bar{c}(G_s)}=0$ if and only if $c(i)=0 \: \forall i \in G_s$, formula (\ref{rel2}) cannot be defined for subgraphs that are characterized only by nodes with a null centrality\footnote{Betweenness centrality vanishes, for instance, for pendant nodes, so the subgraph $G_{s}$, induced by pendant nodes, has an average betweenness equal to zero.}.

Formula (\ref{rel2}) highlights two specific components in the relative centrality of a node with respect to the average behaviour of the network:
\begin{itemize}
	\item $c(i|G_s )$ is the relative incidence of the centrality of a node with respect to the average behaviour of the considered subnetwork to which the node belongs.
	\item $r_{G_s}$ quantifies how much the average centrality of the considered subnetwork is far from the average behaviour of the network.
\end{itemize}

%In this way, through $r_{G_s}$ we are able to assess if a node belongs to a central subnetwork or not, and by means of $c(i|G_s )$ we catch if the node is relevant in its subnetwork or not. For instance, a node could be important by itself, because it is essential in conveying information in the network, but it belongs to a group that is not so important.  \\
In this way, through $c(i|G_s )$ we catch if the node is relevant in its subnetwork or not; by means of $r_{G_s}$ we are able to take into account of the subnetwork position in the whole network. For instance, the disaggregated terms in formula (\ref{rel2}) may suggest that a node could be important by itself, because it is essential in conveying information in the network, but it belongs to a group that on average is not relevant with respect to the whole network.  \\
%XXXX QUESTA PARTE MI SEMBRA PERICOLOSAMENTE SULLA DIFENSIVA E FORSE SPIEGHEREI ANCHE MEGLIO QUALI SONO I PROBLEMI DI EVERETT1999 It is worth noting that the average centrality ${\bar{c}(G_s)}$, that appears in formula (\ref{rel2}), does not represent a measure of group centrality. Indeed, although using the average of individual centralities could be seen as a natural approach to measuring the group centrality, it can lead to a more than one problem, as pointed out in \cite{Everett1999}. OK VISTO COME PROSEGUE IL TESTO NON VA NEMMENO BENE IN QUESTO PUNTO QUESTA FRASE..HO PENSATO DI MODIFICARLA E NELL'INTRODUZIONE HO MESSO UNA PROPOSTA IN ROSSO XXXX
We proceed further disaggregating the factors of the relative centrality measure, to gain more information. Indeed, it could be interesting to measure the importance of a vertex as element of a subgraph, also referring to the centrality of the considered subnetwork. To this end, we introduce an additional component depending on the centrality of the subnetwork $G_s$. % in the expression of the relative centrality $c(i|G)$ in (\ref{rel2}).
We will call this component ${c}(G_s)$ and we rewrite $c(i|G)$ as:
\begin{equation}
\label{rel3}
c(i|G)=\frac{c(i)}{{c(G_s)}}\frac{c(G_s)}{{\bar{c}(G_s)}}\frac{\bar{c}(G_s)}{{\bar{c}(G)}}=\frac{c(i)}{{c(G_s)}}k_{G_s}r_{G_s}
\end{equation}
%XXXX IN QUESTA FORMULA SOPRA, ESPLICITEREI MEGLIO COSA POSSA RAPPRESENTARE $c(G_s)$. CREDO DI CAPIRE CHE RAPPRESENTI UNA CARATTERISTICA DI $G_s$, MA COME SI DEFINISCE? CREDO DOVREMMO SPIEGARE LA DIFFERENZA TRA $c(G_s)$ E $\bar{c}(G_s)$ VEDETE SE VA BENE QUELLO CHE SCRIVO SOTTO - SI VA BENE. LA CENTRALITA' DI GRUPPO E' UN CONCETTO NON AGEVOLE DI FORMULAZIONE, PERCHE' CERCA DI DEFINIRE QUALCOSA DI PIU' DELLA SEMPLICE AGGREGAZIONE ATTRAVERSO LA SOMMA O LA MEDIA.... LA FRASE CHE HO AGGIUNTO IN ROSSO NELL'INTRODUZIONE CERCA  DI SPIEGARE QUESTA COSA E DICO INOLTRE CHE NOI NELLA NOSTRA FORMULA SFRUTTIAMO ENTRAMBE

The term ${c}(G_s)$ is a measure of the group centrality related to $G_s$. Measures of group centrality have been proposed in \cite{Everett1999} for some well-known vertex centralities.
We avoid to give a definition of it in the general case, and refer to next subsection where some specific cases of centrality measure $c$ will be presented. This said, we are implicitly assuming that $G_s$ is such that ${c}(G_s) \neq 0$, so that definition (\ref{rel3}) is well-posed.

The term $k_{G_s}$ quantifies how much the subnetwork centrality is far from the average centrality. An high value of this ratio is typically due to the higher contribute of the external vertices to the centrality measure of the nodes belonging to the group. On other hand, when the structure of the nodes that are outside $G_{s}$ leads to a high vaue of centrality measure, a lower ratio is observed.
%Specifically, in the following, we will focus on betweenness and eigenvector centralities. For these measures, we propose the relative incidence of the centrality of a node with respect to both the average centrality of the subgraph ($\bar{c}(G_s)$) and the centrality ${c}(G_s)$ of the subnetwork to which the node belongs. We assume that the centrality measures we deal with are normalized. It is worth noting that (\ref{rel3}) cannot be formulated when $c(G_s)=0$. This depends, at the end, on the used measure and we will discuss separately these situations.

%XXXX]

\section{Empirical experiments on the market network}
\label{sec:emp}
This section is devoted to the illustration of the usefulness of the relative centrality measure and of its components. As we will see, we present the paradigmatic cases of betweenness and eigenvector centrality applied to financial markets.

\subsection{Description of the dataset and construction of the networks}\label{sec:descr}
In this section, %we perform some empirical applications in order to assess the effectiveness of the proposed approaches.
we test the proposed approaches performing some empirical applications.
We collected daily returns of a dataset referred to the time-period ranging from January 2001 to the end of 2017, that includes $102$ leading U.S. stocks constituents of the $S\&P$ 100 index at the end of 2017\footnote{Data have been downloaded from \cite{bloombe}.}.  Returns have been divided by using monthly stepped two-years windows. %It means that
More precisely, the data of the first in-sample window of width two years are used to build the first network, therefore the process is repeated rolling the window one month forward until the end of the dataset is reached. We obtain a totality of 181 networks, the first one, denoted as \lq\lq1-2001\rq\rq covers the period Jan 1st, 2001 to Dec $31^{nd}$, 2002. The latter one (\lq\lq 1-2016\rq\rq) covers the period  Dec $1^{st}$, 2016 to Dec $31^{nd}$, 2017.\\
As a result, for each window, we have a network $G_{t}=(V_{t},E_{t})$ (with $t=1,...,181$), where nodes are the assets and edges are weighted by computing the correlation coefficient ${}_{t}\rho_{i,j}$ between each couple of assets. Notice that the number of assets can vary over time. We have indeed considered the 102 assets constituents of the $S\&P$ 100 index at the end of 2017. For some of these assets no information are available in some specific time periods. Therefore, in each window we have considered only assets whose observations are sufficiently large
%XXX QUALE E' LA SOGLIA DEL SUFFICIENTLY LARGE? DA AGGIUNGERE SOGLIA
%L HO MESSA IN NOTA NON VOLEVO ESSERE TROPPO ESPLICITO. PURTROPPO NON ABBIAMO POSSIBILITA' di MODIFICA SU QUESTA COSA IN QUANTO I DATI USATI SONO FRUTTO DI UNA PREELABORAZIONE, FATTA IN UN ALTRO PAPER E NON DA ME E ROSANNA) . IN OGNI CASO IL NUMERO DI NA CHE USIAMO PER RIMUOVERE GLI EVENTUALI ASSETS E' MOLTO PRUDENZIALE ...
to assure a significant estimation of the correlation coefficient\footnote{In the empirical application, in a window $t$ we disregard assets with a number of missing data higher than 20.}. As a consequence, the number of nodes in the 181 networks varies from 83 to 102 during the time-period. \\
%As well known, correlation network belongs to a class of networks based on similarity, where a weighted edge indicates an affinity (but not necessarily a direct interaction) between the two nodes. Despite a variety of the similarity measures, a widespread choice consists in quantifying the similarity between two elements (nodes) of the system with the Pearson coefficient. Furthermore, it is worth pointing out that considering all correlations between assets, we obtain a complete structure. \\
%Some alternative approaches have been provided in literature, in order to reduce the complexity of the network. For instance, suitable techniques can be used to filter the original network, in order to study the centrality the assets. In particular, a well-known filtering method is based on hierarchical method, as that consisting in extracting the minimum spanning tree (MST) (see \cite{Onnela_2003} and \cite{Mantegna}) or on the planar maximum filtered graph (PMFG) (see \cite{pozzi}). These are devised to specifically detect  the hierarchical structure of the data, namely, the elements of the system can be partitioned into clusters which, in turn, can be partitioned into subclusters, and so on up to a certain level. \\
%Another approach is based on constructing a threshold network, where all correlation coefficients lower than a given threshold are discarded.
As already mentioned in the introduction, in the present analysis we filter $G_{t}$ ($t=1,..,181$) considering only the edges whose associated correlation coefficients are larger than $0.3$ (i.e. we obtain, for each time period, a network $G^{F}_{t}=(V_{t},E^{F}_{t})$). This value has been estimated as suggested in \cite{battiston2010} and the approach can be useful to preserve only links associated with statistically significant correlation. \\
Since the analysis of assets centrality seems a relevant topic in the related literature, by referring to the filtered networks we focus here on the study of the relative importance of an asset with respect to the portfolio of all assets as well as those characterized by assets of the same sector.

\subsection{Employed relative centrality measures}
We now introduce the relative centrality measures employed in the analysis. Their formalization mirrors the general arguments of Section \ref{sec:c}, with some details that are reported for the sake of clarity.

\subsubsection{Relative Betweenness Centrality}
\label{RBet}

The shortest-path betweenness centrality (\cite{Freeman79}) quantifies how often a node is located on a shortest path between all other nodes. Formally, it is the percentage of geodesics between pairs of vertices $j,k\neq i$, passing through $i$:
\begin{equation}
b(i)=\sum_{j<k}\frac{g_{jk}\left(  i\right)  }{g_{jk}}\label{eq:bet}%
\end{equation}
where $g_{jk}$ is the number of geodesics from node $j$ to node $k$, and $g_{jk}\left(
i\right)  $ is the number of those geodesics that pass through $i$. The normalized measure is $\frac{b(i)}{{n-1 \choose 2}}$.

Formula (\ref{rel}), applied to the specific case of betweenness centrality, becomes:
\begin{equation}
	b(i|G)=\frac{b(i)}{{\bar{b}(G)}}
	\label{eq:relbet1}
\end{equation}
where ${\bar{b}(G)}=\frac{\sum_{i=1}^nb(i)}{n}$ is the average betweenness of $G$. \\
Let us suppose that $i$ belongs to a subgraph $G_{s}$; then formula (\ref{rel2}) is, in this case:
\begin{equation}
\label{eq:scompbet1}
	b(i|G)=\frac{b(i)}{{\bar{b}(G_s)}}\frac{\bar{b}(G_s)}{{\bar{b}(G)}}=b(i|G_s )r_{G_s}
\end{equation}
where ${\bar{b}(G_s)}=\frac{\sum_{i \in G_s}b(i)}{n_s}$ is the average betweenness of $G_s$.
%The interpretation of the components $b(i|G_s)$ and $r_{G_s}$ follows accordingly to the previous Section.

According to the general concept in Section \ref{sec:c}, we intend to quantify the intermediary role position of vertex $i$ taking into account also of the centrality of the subnetwork $G_s$. As previously said for centrality in general, a measure of betweenness centrality referred to a subset of vertices in a network (the so-called \emph{group betweenness centrality}) has been introduced by \cite{Everett1999} in a more general context.
For convenience of the reader, we remind here the definition. $\forall j,k \in G \setminus G_s$, let $g_{jk} (G_s)$ be the number of $j-k$ geodesic paths passing through at least one vertex of $G_s$.
The group betweenness centrality of $G_s$ is\footnote{The normalized group betweenness can be obtained by dividing each value by the theoretical maximum, yielding to $b'(G_s)= \frac{2b(G_s)}{(n-n_s)(n-n_s-1)}.$ }:
\begin{equation}
\label{Gbet}
b(G_s)=\sum_{j<k}\frac{g_{jk}\left(  G_s\right)}{g_{jk}}, \quad j,k \in G \setminus G_s
\end{equation}
Group betweenness measures the betweenness of $G_s$ only referring to
the paths leading to the external vertices, i.e. vertices that do not belong to the subgraph.

According to formula (\ref{rel3}), $b(i|G)$ can be rewritten as:

\begin{equation}
b(i|G)=\frac{b(i)}{{b(G_s)}}\frac{b(G_s)}{{\bar{b}(G_s)}}\frac{\bar{b}(G_s)}{{\bar{b}(G)}}=b^{G}(i|G_s)k^{b}_{G_s}r^{b}_{G_s}
\label{eq:scompbet2}
\end{equation}

Through the previous formula the relative betweenness of a node can be seen with respect to the average behaviour of the network, in three components:
\begin{itemize}
	\item $b^G(i|G_s)$ measures how much the node $i$ is essential in conveying information with respect to the intermediary role of its subnetwork;
	\item $k^{b}_{G_s}$ quantifies how much the betweenness of the subnetwork is far from the average betweeness. An high value of this ratio is achieved in presence of high contribution to the betweenness of the nodes of $G_s$ of the nodes outside $G_s$, which means that $G_s$ is relevant for conveying information among nodes not belonging to $G_s$.
	\item $r^{b}_{G_s}$ quantifies how much the average betweenness of $G_s$ is far from the average betweenness of the entire network, hence measuring the discrepancy between $G_s$ and $G$ in terms of inner connectivity.
\end{itemize}

Notice that, the group betweenness centrality definition provided by formula (\ref{Gbet}) allows $b(G_s)=0$. We are implicitly assuming that $b(G_s) \neq 0$ since formula (\ref{eq:scompbet2}) is meaningless otherwise; however, it could be interesting to analyse also the case of $b(G_s)=0$, to intercept extremal situations. Indeed, some individuals could have a non-zero betweenness centrality although they are member of a subnetwork with zero group betweenness. In this case, we can measure $b(i|G)$ by using formulas (\ref{eq:relbet1}) and (\ref{eq:scompbet1}), but obviously it does not make sense to evaluate the component $b^G(i|G_s)$.

\subsubsection{Relative Eigenvector Centrality}
\label{REI}

The eigenvector centrality is an extremely important measure of vertex influence in the network. The meaning of this measure stems from the fact that a vertex is highly central if it is adjacent to vertices that are themselves highly central. \\
In a formal way, the centrality score is defined using the Perron vector $\mathbf{x}$.
%Let $\mathbf{A}$ be the adjacency matrix associated with the graph $G=(V,E)$, $\rho$ the spectral radius and $\mathbf{x}$ the principal eigenvector. It is well known from matrix theory that all components of $\mathbf{x}$ are real and strictly positive.
More precisely, the eigenvector centrality
(\cite{Bonacich1972,Bonacich1987}) is defined as:
\begin{equation}
x(i) =\frac{1}{\rho}\sum_{j=1}^na_{ij}x(j),
\end{equation}
where $\rho$ is the spectral radius of the adjacency matrix $\mathbf{A}$, as explained in Section \ref{sec:prelnot}.
In this way, not only the number of adjacent nodes contributes to the node centrality, but also their centralities. Since the node centrality is reinforced by the centralities of its neighbours, this measure well captures the power of a vertex in a network. %Using the matrix expression $\mathbf{x}=\frac{1}{\rho}\mathbf{Ax}$.
The normalized eigenvector measure is $\frac{\mathbf{x}}{\parallel \mathbf{x} \parallel_{\tiny 2}}$, where $\parallel\mathbf{x} \parallel_{\tiny 2}$ is the Euclidean norm.

Focusing on the eigenvector, formula (\ref{rel}) becomes:

\begin{equation}
x(i|G)=\frac{x(i)}{{\bar{x}(G)}}
\label{releigen}	
\end{equation}

where ${\bar{x}(G)}=\frac{\sum_{i \in G}x(i)}{n}$ is the average eigenvector centrality of $G$. Notice that ${\bar{x}(G)}$ never vanishes, since $x(i)\neq0 \: \forall i \in G$. \\
Formula (\ref{rel2}), that highlights the relative centrality with respect to the average centrality of the subgraph, becomes in this case:

\begin{equation}
x(i|G)=\frac{x(i)}{{\bar{x}(G_s)}}\frac{\bar{x}(G_s)}{{\bar{x}(G)}}=x(i|G_s )r^{x}_{G_s}
\label{releigen2}
\end{equation}

where ${\bar{x}(G_s)}=\frac{\sum_{i \in G_s}x(i)}{n_s}$ is the average eigenvector centrality of $G_s$. %Once again, the interpretation of the components $x(i|G_s)$ and $r^{x}_{G_s}$ follows accordingly to Section (\ref{RCent}).\\

As already done in Section (\ref{RBet}), we want to provide a measure of the eigenvector centrality of the subnetwork, using a measure of group centrality.
The idea is to replace all the nodes of the subnetwork by a single node whose neighbourhood is the union of the neighbourhoods of all subnetwork members. In other words, an edge from the new vertex to another one exists if there is at least one vertex in the subnetwork who had that link. Through this approach, named in the literature \lq\lq Reduced Model Approach'', we generate a new graph $G^*$ (\textit{reduced graph}) of $n-n_s+1$ vertices, of which we can compute the individual centralities in order to obtain the centrality measure for the subset\footnote{There exist other approaches in the literature to compute the centrality of a subset, such as, for instance, those proposed in \cite{Bonacich1991}.}. Using the Reduced Model Approach we can then compute the eigenvector centrality $x(G_s)$ referred to the subnetwork $G_s$.
Hence, formula (\ref{rel3}) is in this case equal to:
\begin{equation}
\label{releigen3}
x(i|G)=\frac{x(i)}{{x(G_s)}}\frac{x(G_s)}{{\bar{x}(G_s)}}\frac{\bar{x}(G_s)}{{\bar{x}(G)}}=x^G(i|G_s)k^{x}_{G_s}r^{x}_{G_s}
\end{equation}

Moving to the interpretation, we then extrapolate from (\ref{releigen3}) the following components:
\begin{itemize}
	\item $x^G(i|G_s)$ relates the power/influence of the node $i$ with respect to the power of its subnetwork. In other words, it measures the individual power in respect to the collective power. This component gives insights about the fact that the node is powerful \lq\lq by himself'' or its power arises from its group membership;
	\item $k^{x}_{G_s}$ quantifies how much the subnetwork is powerful with respect to the average power;
	\item $r^{x}_{G_s}$ quantifies how much the subnetwork is powerful on average with respect to the entire network.
\end{itemize}

\noindent Notice that, unless the group betweenness measure, the group eigenvector is always greater than zero, given the connectivity assumption on the network $G$. 

\subsubsection{Some remarks on the selected centrality measures}
\label{remarks}

Further remarks can be made about the choice of the most appropriate centrality measure, namely, the measure that better identifies the idea of being ``influential''. As pointed out in the introduction, eigenvector centrality captures influences at long distances. More precisely, whereas degree centrality measures the local influence of a node, the eigenvector centrality captures the global influence.
%\added{Some further remarks can be done about the choice of the most appropriate centrality measure, meaning the measure that better indentifies the idea of be ``influential''. As pointed out in the introduction, eigenvector centrality captures influences at long distances. More precisely, whereas degree centrality measures the local influence of a node, the eigenvector centrality measures the global influence of the node.}
However, although widely used, eigenvector centrality also presents some limitations. Depending on the network structure, most of the weights of the eigenvector could be concentrated in few nodes, like hubs.  In this case, most of the nodes will present centrality close to zero and, therefore, the importance of nodes is not well quantified.  For instance, \cite{Martin2014} show that, in random networks with only one high-degree hub or power-law degree distributions, the leading eigenvector can undergo a localization in which most of the weight of the vector is concentrated around the hub vertex and its neighbours, whereas the centrality of the remaining nodes vanishes for large networks\footnote{ \cite{Martin2014} overcome this issue providing a new centrality measure based on the leading eigenvector of the Hashimoto or non-backtracking matrix}.
Additionally, \cite{Landherr2010} show that eigenvector centrality does not display perfect monotonicity with respect to distance and shortest path. However, an empirical analysis, conducted on the robustness of measures of centrality in the face of random error in the network data, show that four different centrality measures (betweenness, closeness, degree and eigenvector centrality) are surprisingly similar with respect to pattern and level of robustness in random networks (see \cite{Borg_2006}). \\
Other centrality measures, in line with this one, have been provided in the literature.  Among them, Katz centrality (see \cite{Katz}) takes into account short, medium and long range influences, modulated by an attenuation factor $\alpha$.
%In this framework, our proposal can be adapted also to other centrality measures. For instance, unlike eigenvector centrality that emphasizes long distance influences, Katz centrality (see \cite{Katz}) try to take into account short, medium and long range influences. 
Formally, it is defined as row sums of the matrix $(\mathbf{I}-\alpha\mathbf{A}^{-1})$, that is the sum of the series of $\alpha^k\mathbf{A}^k$, being $0 < \alpha < \lambda_1 $, where $\lambda_1$ is the largest eigenvalue of $\mathbf{A}$. 
According to Katz, not only the number of direct connections but also the further interconnectedness of nodes plays  an  important  role  for  the  overall  interconnectedness  in  a  social  network. Therefore, Katz includes all walks of arbitrary length from the considered node to the other nodes of the network, penalizing the contribution of walks of length $k$ by $\alpha^k$. Hence, this centrality measure falls in the middle between the local measure (the degree) and the global one (the eigenvector). The selection of the attenuation factor adds another challenge. Different choices of $\alpha$ lead to different node rankings (\cite{Benzi2015}).  \\
However, it is worth pointing out that formula (\ref{releigen3}) can be also provided for the case of Katz centrality. Indeed, it is possible to compute the Katz centrality of the subnetwork $G_{s}$ by applying the Reduced Model Approach previously described.
%\color{blue}{This fact put the measure in the middle between the local measure (the degree) and the global (the eigenvector). 
%The attenuation factor is important, but its choice adds a further difficulty  respect to the other considered measures, especially because different choices can produce different scores leading to different node rankings (\cite{Benzi2014}).  
%Formula (\ref{releigen3}) can be provided also for the case of Katz centrality. Indeed, it is possibile to compute the Katz centrality of the subnetwork $G_{s}$ by applying the Reduced Model Approach previously described.}

\section{Results and Discussion}\label{sec:res}
Given the filtered network $G^{F}_{t}=(V_{t},E^{F}_{t})$ (with $t=1,...,181$), derived as described in Section \ref{sec:descr}, we initially computed betweenness and eigenvector centralities, that have been explored by previous works in this field (\cite{pozzi}, \cite{Peralta}). In particular, by means of formula (\ref{rel}), the relative incidence for each node and for each measure have been obtained.

\begin{figure}[!h]
	\includegraphics[scale=0.5]{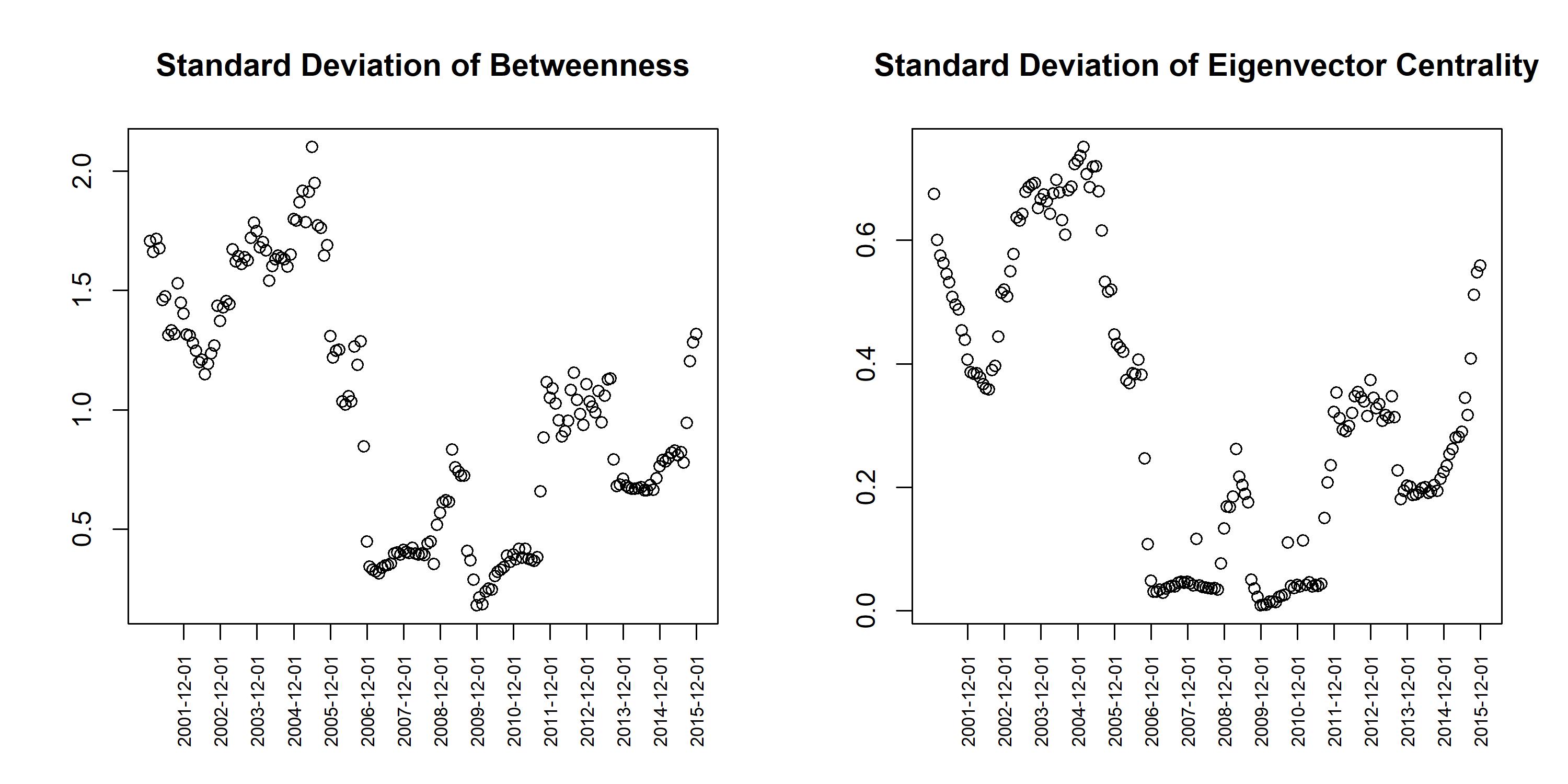}
	\caption{Given the specific filtered network $G^{F}_t$ (with $t=1,...,181$), the relative betweenness centrality $b(i|G^{F}_{t})$ of each node $i$ has been computed. The process is repeated for each time period and, for each $t$, we compute the standard deviation of the distribution of the relative centrality measures obtained. Results are displayed on the left side. On the right side, the same procedure has been applied by considering the relative eigenvector centrality $x(i|G^{F}_{t})$.} 
	\label{fig:sd}
\end{figure}

At the global level, an interesting result is provided by the behaviour of the standard deviation of the relative nodes' centrality over time (see Figure \ref{fig:sd}). Although slight differences, we observe that the standard deviation of both relative centrality measures tends to decrease in period of crisis. Both the financial crisis period in 2007-2008, which is identified with the Lehman Brothers failure, and the Sovereign debt crisis in 2010-2011 are noticeable. This behaviour reflects the fact that the correlation between assets is higher in this period, leading to an increase in the density of the filtered network and then to similar behaviours in terms of centrality between assets. Such an outcome meets the well-known stylized fact in finance, for which assets correlation increases in times of financial distress.

\begin{figure}[!h]
	\includegraphics[scale=0.4]{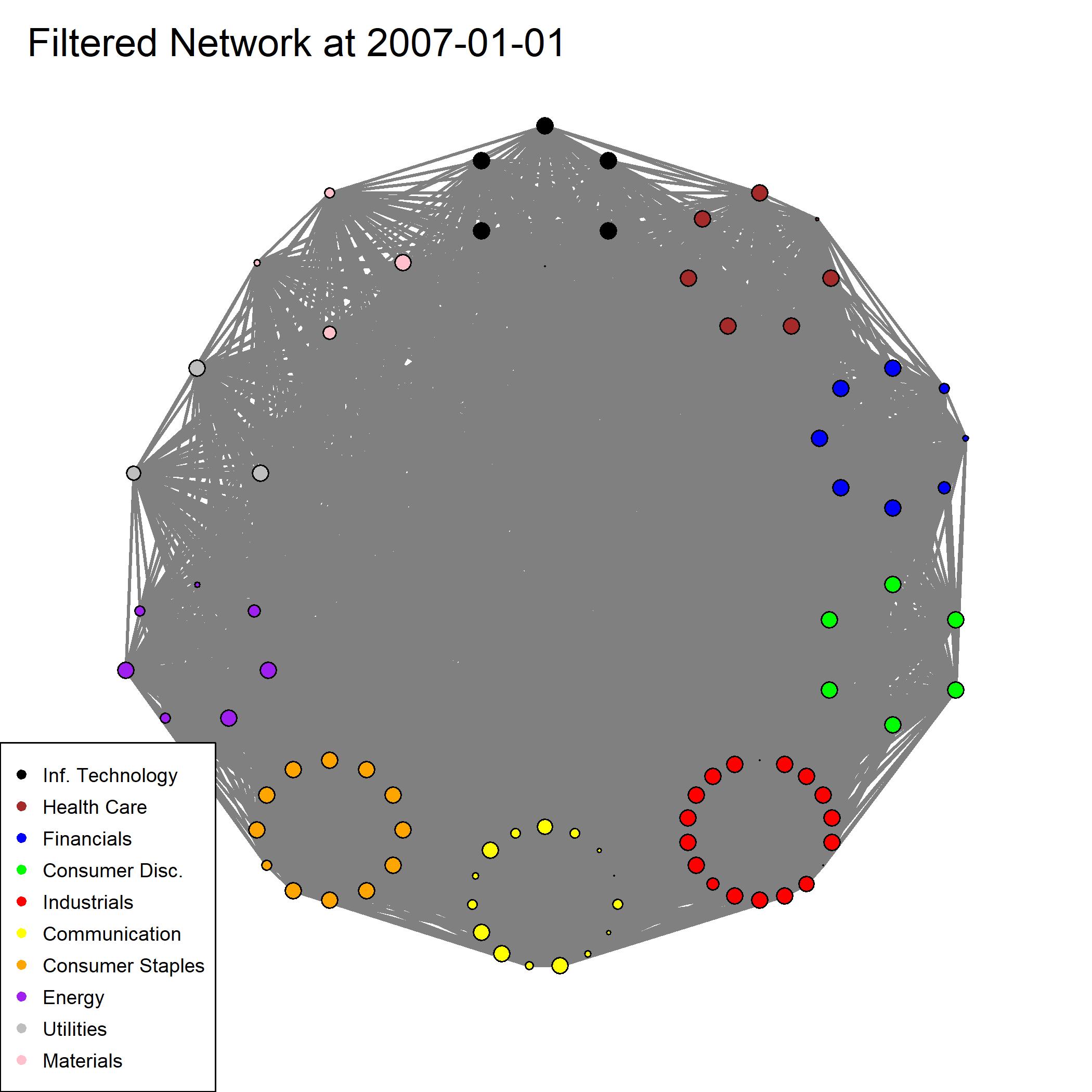}
	\includegraphics[scale=0.4]{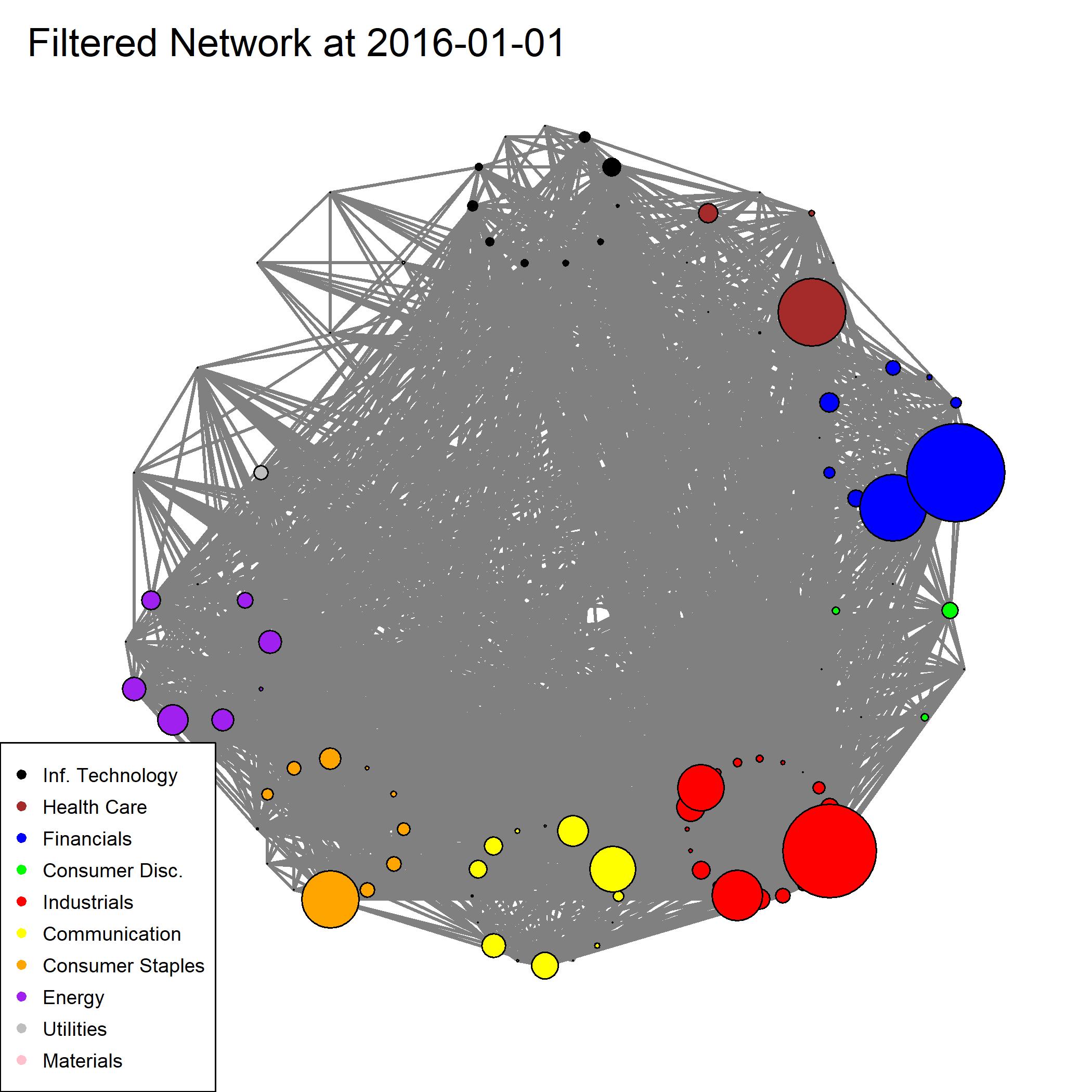}
	\includegraphics[scale=0.4]{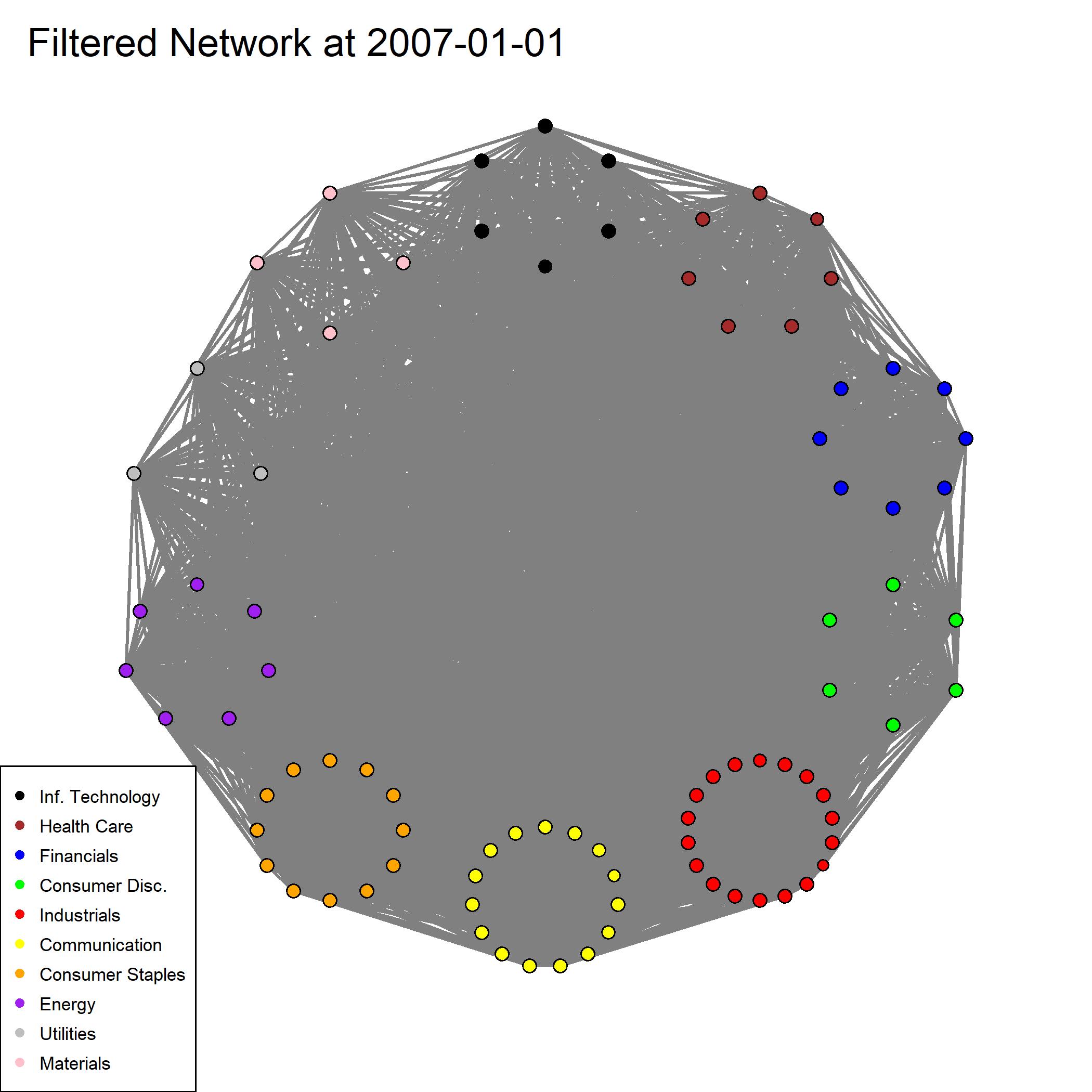}
	\includegraphics[scale=0.4]{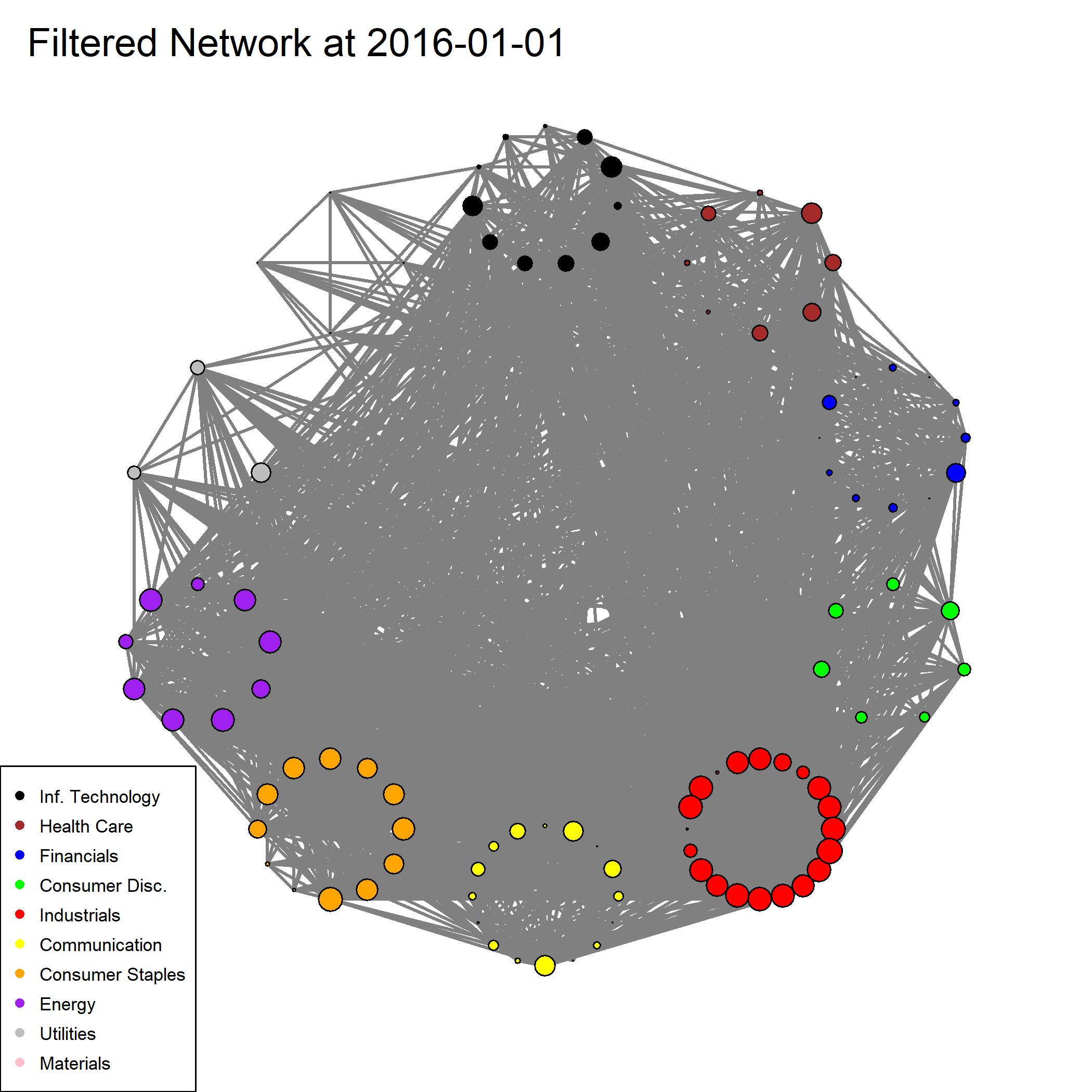}
	\caption{Filtered Networks at the end of 2006 and 2015. \added{They refer to data of the two-year periods 2007-2008 and 2016-2017, respectively.  Nodes are assets and edges' weights are related to the correlation coefficients between returns of couple of assets (see Section \ref{sec:descr} for details).} \deleted{referred to the two centrality measures, namely, betweenness and eigenvector centrality.} Assets are grouped in 10 sectors, according to the financial classification \added{reported in the legend}. On the upper side, \added{we focus on the role of relative betweenness centrality of each asset, namely,} the bullets size is proportional to $b(i|G^{F}_{t})$. On the lower side, the bullets size is proportional to the relative eigenvector centrality.}
	\label{fig:Net}
\end{figure}

%[XXXXX COME DICEVO PRIMA, NON HO CHIARISSIME QUESTE DUE RETI
Concerning specific assets, we report in Figure \ref{fig:Net} the networks $G^{F}_{t}$  \lq\lq1-2007\rq\rq and \lq\lq1-2016\rq\rq. They refer to data of the two-year periods 2007-2008 and 2016-2017, respectively. Assets have been classified in 10 sectors, according to the standard sector classification defined by the Global Industry Classification Standard developed by Morgan Stanley Capital International and Standard \& Poor’s\footnote{For a detailed description of sectors see, for instance, Appendix 1 in \cite{Beber2011}}. In Figure \ref{fig:Net} we relate the size of the nodes to the value of the relative centrality (betweenness for the upper figures and eigenvector for the lower ones). As already stressed, we observe both a higher average centrality and higher differences between assets in quiet periods. Some sectors appear prominent in terms of centrality, also showing a greater homogeneity between nodes. On the contrary, other sectors show a significant heterogeneity, with a few nodes extremely central and several non-central ones (see, for instance, Consumer Staples  and Health Care in the \lq\lq 1-2016\rq\rq network based on relative betweenness centrality). 

\begin{figure}[!h]
	\includegraphics[scale=0.4]{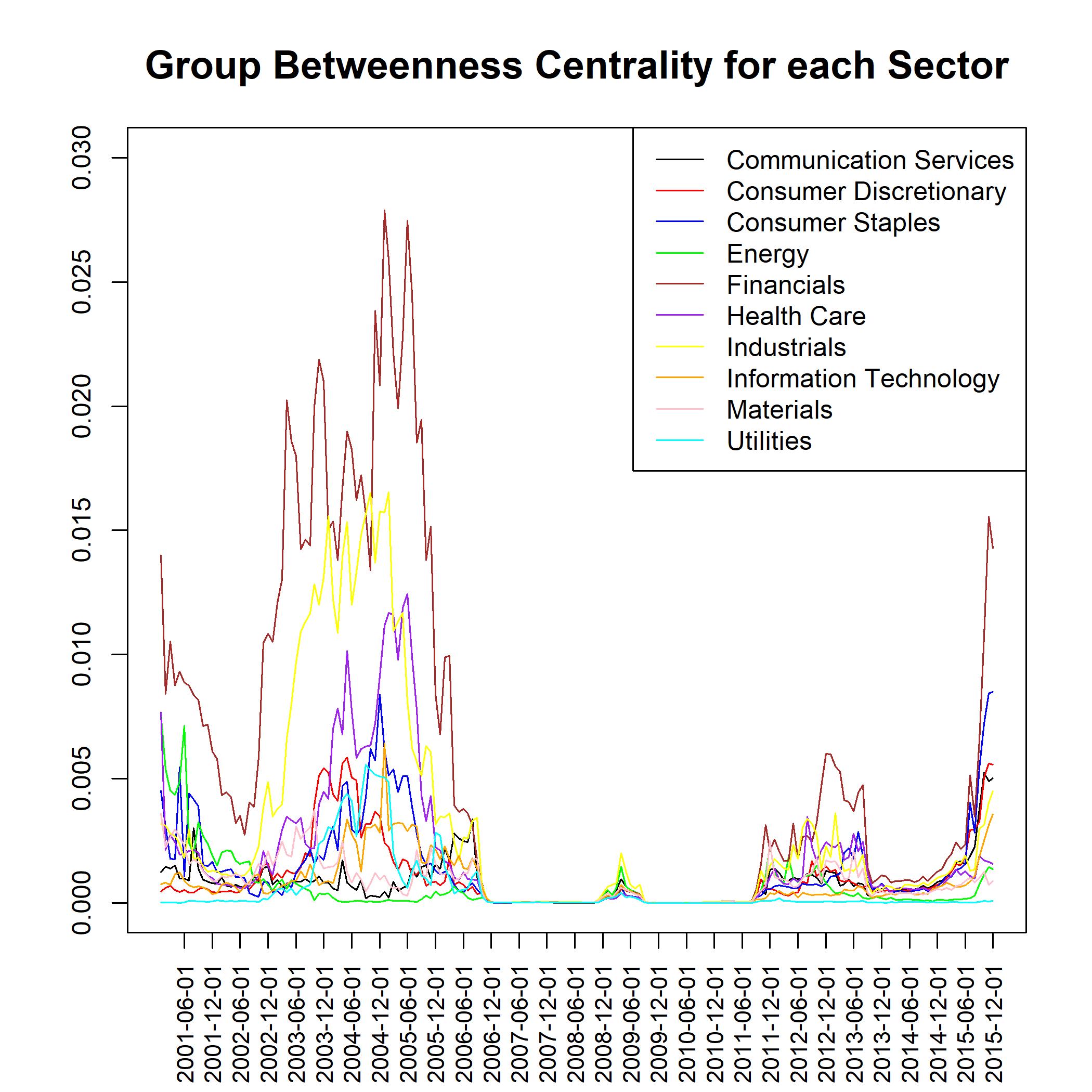}
	\includegraphics[scale=0.4]{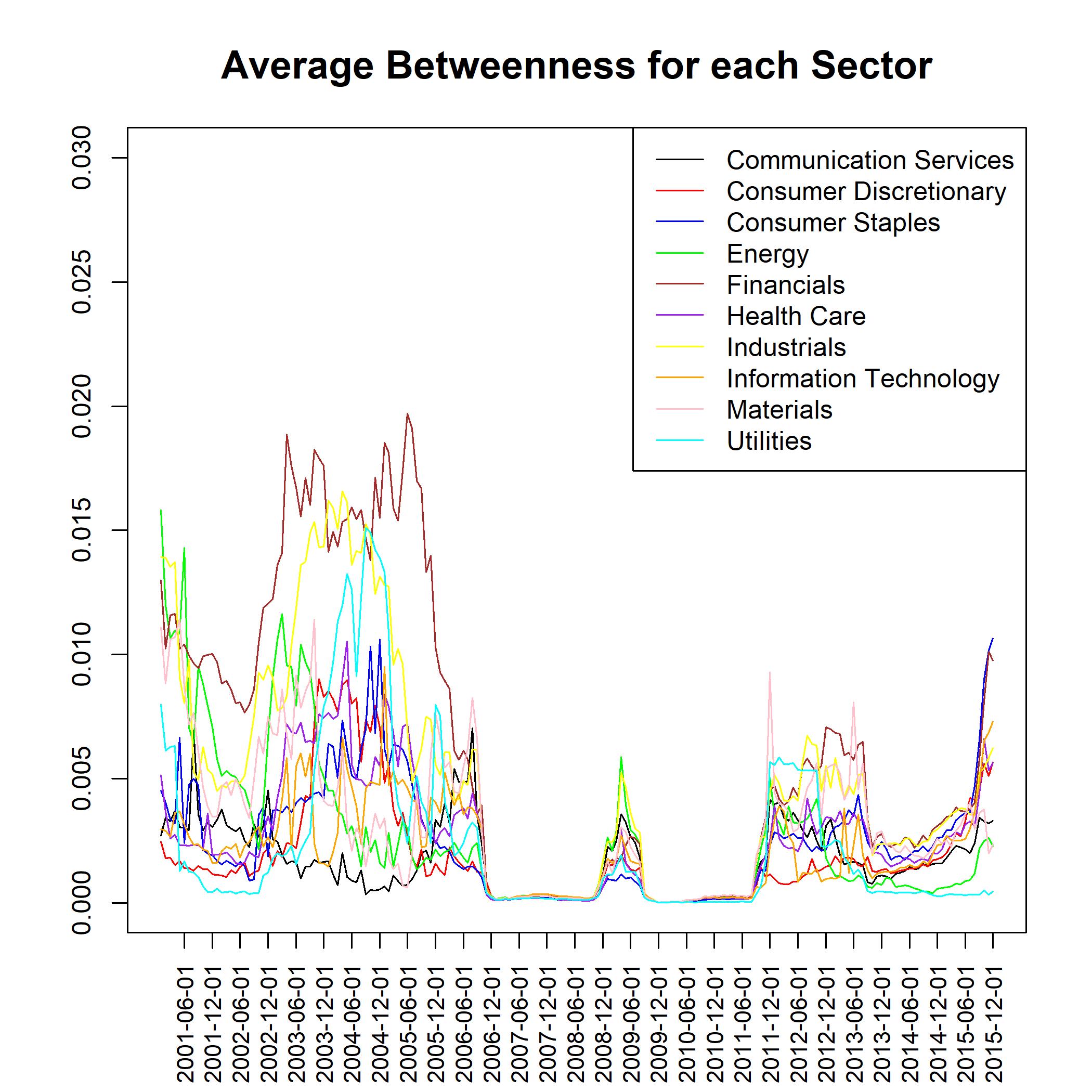}
	\caption{\added{We display, on the left side,} group betweenness centrality $b(G_{s})$ \added{for each sector $s=1,...,10$ and for different time periods $t=1,...,181$.} The average betweenness centrality $\bar{b}(G_{s})$ for each sector $s=1,...,10$ and for different time periods $t=1,...,181$ \added{is instead reported on the right side.} }
	\label{fig:SecNetBet}
\end{figure}

This preliminary analysis suggests that it could be interesting to investigate not only the centrality of an asset with respect to both the financial market and the sector to which the asset belongs, but also the role/position of each sector in the whole network. According to formula (\ref{eq:scompbet2}), we computed both the average betweenness and the group betweenness centrality (see Figure \ref{fig:SecNetBet}). It is interesting to note that both indicators show a similar pattern over time, highlighting again differences between quiet and more turbulent periods. In both cases, we observe that, on average, the \emph{Financial}, \emph{Industrial} and \emph{Consumer Staples} sectors are the most central. The \emph{Information Technology} sector, that is actually the prominent sector of $S\&P$ 100
%XXX sempre stato S\&P100 XXX
in terms of market capitalization, is increasing its centrality over time. It shows indeed a very low centrality with respect to other sectors at the beginning of the period and it has slowly increased his ranking over the last decade. On the other hand, the \emph{Energy} subnetwork is extremely central in 2001, while it shows a very low centrality since 2012.

Now, we focus on the behaviour of specific assets and we report in Tables \ref{tab:1} and \ref{tab:2} both the top and the bottom rankings in terms of relative betweenness centrality (computed by using formula (\ref{eq:relbet1})). The three ratios of the decomposition provided by formula (\ref{eq:scompbet2}) are also displayed. Values have been computed by considering the last period available (i.e. network  \lq\lq1-2016\rq\rq). We observe that the most central assets do not necessarily belong to the most central sectors. On one hand, the financial sector is strongly represented by 6 assets in the top 15 ones. These assets are central in the network, but many of them are, at the same time, also relevant in their sector (as shown by the value of $b^{G}(i|G_{s})$). On the other hand, we can notice some examples of assets (like Home Depot and Amgen, for instance) that are really relevant in their sector, but belongs to a sector that, on average, is not very representative at the global level (as the $r^{b}_{G_{s}}$ coefficient shows for the Health Care and the Consumer Discretionary sectors). \\
It is also interesting to note that Amazon, that is one of the top ten constituents of the $S\&P$ 100 index\footnote{See Standard and Poor Factsheets, 2019}, has a definitely low centrality in the network and it is also not extremely relevant in its sector. \\
Furthermore, we analyse in Table \ref{tab:3} how the top centrality assets are changed over time. To this end, we compare the top ranking in terms of relative betweenness in periods before and after the two crises, respectively.  We observe how the composition drastically changed over time. At the beginning of the period (namely, the two-year 2001-2002), the financial and energy sectors almost make up completely the top group.  The composition is a bit different in the subsequent periods, where assets of various sectors increase their central role. Interesting situation occurs immediately after the financial crisis, where industrial, costumer staples and health care appear as the prominent sectors, while the financial sector, except for some specific cases, significantly reduces his centrality. As noticed before, over the last period (namely the two-year 2016-2017), the financial sector becomes again prevalent.

\begin{table}[!h]
	\begin{tabular}{|lc|c:ccc|}
		\hline\hline
		Asset Name & Sector & $b(i|G)$ & $b^{G}(i|G_s)$ & $k^{b}_{G_s}$ & $r^{b}_{G_s}$ \\
		\hline\hline
		Mondelez & Consumer Staples	& 7.04	& 5.37	& 0.80	& 1.65 \\
		Berkshire Hathaway	&  Financials	& 6.71	&3.05	&1.46	&1.51 \\
		Home Depot	& Consumer Discretionary	& 4.84	&5.63	&0.98	&0.88 \\
		PepsiCo Inc	& Consumer Staples	& 4.77	&3.64	&0.80	&1.65 \\
		Honeywell International	& Industrials	& 4.12	&5.93	&0.72	&0.96 \\
		Mastercard	& Financials	& 3.63	&1.65	&1.46	&1.51 \\
		Visa	& Financials	& 3.34	&1.51	&1.46	&1.51 \\
		Amgen	& Health Care	& 3.29	&13.45	&0.28	&0.87 \\
		Blackrock	& Financials	& 2.82	&1.28	&1.46	&1.51 \\
		Abbott Laboratories	& Health Care	& 2.21	&9.06	&0.28	&0.87 \\
		Microsoft Corp	& Information Technology	& 2.19	&3.96	&0.49	&1.13 \\
		Coca-Cola Company	& Consumer Staples	& 2.04	&1.56	&0.80	&1.65 \\
		U.S. Bancorp	& Financials	& 2.04	&0.93	&1.46	&1.51 \\
		Danaher Corp	& Health Care	& 1.91	&7.81	&0.28	&0.87 \\
		Johnson \& Johnson	& Health Care	& 1.76	&7.19	&0.28	&0.87 \\
		\hline\hline
	\end{tabular}
	\caption{Top 15 ranking in terms of relative betweenness based on the network \lq\lq 1-2016\rq\rq}
	\label{tab:1}
\end{table}

\begin{table}[!h]
	\begin{tabular}{|lc|c:ccc|}
		\hline\hline
		Asset Name & Sector & $b(i|G)$ & $b^{G}(i|G_s)$ & $k^{b}_{G_s}$ & $r^{b}_{G_s}$ \\
		\hline\hline
		
		Amazon.com &	Consumer Discretionary	&0.025	&0.029	&0.982	&0.875 \\
		ConocoPhillips	&Energy	                &0.022	&0.104	&0.593	&0.353 \\
		Raytheon Company	&Industrials	    &0.021	&0.030	&0.721	&0.964\\
		Occidental Petroleum 	&Energy	    &0.019	&0.090	&0.593	&0.353\\
		American Express Company	&Financials &0.018	&0.008	&1.464	&1.505\\
		Duke Energy 	&Utilities	        &0.017	&1.186	&0.194	&0.072\\
		Charter Communications	&Communication Services	&0.015	&0.019	&1.521	&0.510\\
		Nike	&Consumer Discretionary	        &0.014	&0.017	&0.982	&0.875\\
		Monsanto	&Materials	                &0.012	&0.091	&0.372	&0.366\\
		Lockheed Martin 	&Industrials	&0.012	&0.017	&0.721	&0.964\\
		Southern Company	&Utilities	        &0.009	&0.644	&0.194	&0.072\\
		Bristol-Myers Squibb Company	&Health Care	&0.006	&0.024	&0.280	&0.872\\
		Time Warner	&Communication Services	    &0.006	&0.007	&1.521	&0.510\\
		CVS Health Corp	&Health Care	        &0.002	&0.008	&0.280	&0.872\\
		Allergan Plc	&Health Care	        &0.000	&0.000	&0.280	&0.872\\
		\hline\hline
	\end{tabular}
	\caption{Bottom 15 ranking in terms of relative betweenness based on the network \lq\lq 1-2016\rq\rq}
	\label{tab:2}
\end{table}

\begin{landscape}
	\topskip0pt
	\vspace*{\fill}
   \begin{table}[htb!]
	\centering
	\footnotesize
	\setlength{\tabcolsep}{2pt}	\begin{tabular}{|l c c :l c c: l c c : l c c  |}
		\hline\hline
		\multicolumn{3}{|c:}{2001-2002} &		\multicolumn{3}{:c:}{2004-2005} & 		\multicolumn{3}{:c:}{2013-2014} 	&	\multicolumn{3}{:c|}{2016-2017} \\

		Asset & Sector & $b(i|G)$ & Asset & Sector & $b(i|G)$ & Asset & Sector & $b(i|G)$ & Asset & Sector & $b(i|G)$ \\
		\hline\hline
		American Express &	FI	& 7.11	& Danaher	& HC	& 8.24	& Berkshire Hathaway & FI	& 3.55 &	Mondelez & CS &7.04 \\
		Citigroup & FI	& 7.03 & 	JP Morgan & FI	& 7.34 & 	Johnson \& Johnson	&  HC & 	3.36 & 	Berkshire Hathaway & 	FI	&  6.71 \\
		Morgan Stanley	&  FI	& 4.45	& General Electric & IN & 6.79	& 3M & 	MA & 	2.43 & 	Home Depot	& CD	& 4.84 \\
		Exxon Mobil  &  EN	& 4.27 & Coca-Cola & CS & 6.74	& CVS & HC & 2.23 & PepsiCo & CS	& 4.77 \\
		Goldman Sachs & FI & 	3.55	& Goldman Sachs & 	FI	& 5.01	& Colgate-Palmolive & CS	& 2.22 & 	Honeywell & IN	& 4.12 \\
		United Technologies & IN & 	3.21 & 	Emerson Electric & 	IN & 	4.22 & 	Mondelez & 	CS	& 2,18 & 	Mastercard & 	FI	& 3.63\\
		Bank of New York & 	FI	& 3.14 & 	Home Depot& 	CD	& 3.90 & 	Honeywell & 	IN	& 1.85	& Visa & 	FI& 	3.34 \\
		JP Morgan & 	FI	& 2.95 & 	Southern Company & 	UT& 	3.75 & Danaher	& HC	& 1.85 & 	Amgen & 	HC	& 3.29 \\
		Wells Fargo & 	FI	& 2.94 & 	Bank of America &	FI & 	3,74 & 	General Dynamics & 	IN	& 1.85	& Blackrock & 	FI	& 2.82\\
		Chevron & 	EN	& 2.55 & 	Qualcomm & 	IT	& 2.86 & 	Wells Fargo &	FI	& 1.85	& Abbott Laboratories& 	HC	& 2.21 \\
		General Electric & 	IN & 	2.20 &	Wells Fargo & FI& 	2.70	& PepsiCo & 	CS & 	1.80	& Microsoft & 	IT & 2.19\\
		Walt Disney & 	CO & 	2.14 & 	Caterpillar & 	IN	& 2.49 & Merck & HC & 	1.75 & 	Coca-Cola & CS & 	2.04\\
		U.S. Bancorp & 	FI	& 1.98 & Bank of New York & 	FI& 	2.40 & 	Philip Morris & 	CS & 1.74 & 	U.S. Bancorp & 	FI	& 2.04\\
		Bank of America & 	FI & 	1.78 & 	Morgan Stanley & 	FI & 	2.30 & 	United Technologies & IN	&  1.72 & 	Danaher Corp	& HC& 	1.91\\
		American International Group	& FI	& 1.78 & 	United Technologies & IN & 	2.09	& Union Pacific & 	IN	& 1.71	& Johnson \& Johnson	& HC & 	1.76\\
		
		\hline\hline
	\end{tabular}
	\caption{Top 15 ranking in terms of relative betweenness at different time periods. Sectors have been denoted as follows: CO - Communication Services, CD - Consumer Discretionary, CS - Consumer Staples, EN- Energy, FI - Financial, HC - Health Care, IN- Industrials, IT - Information Technology, MA - Materials, UT - Utilities}
	\label{tab:3}
    \end{table}
\vspace*{\fill}
\end{landscape}%\begin{acknowledgements}
%If you'd like to thank anyone, place your comments here
%and remove the percent signs.
%\end{acknowledgements}

\begin{figure}[!h]
	\includegraphics[scale=0.4]{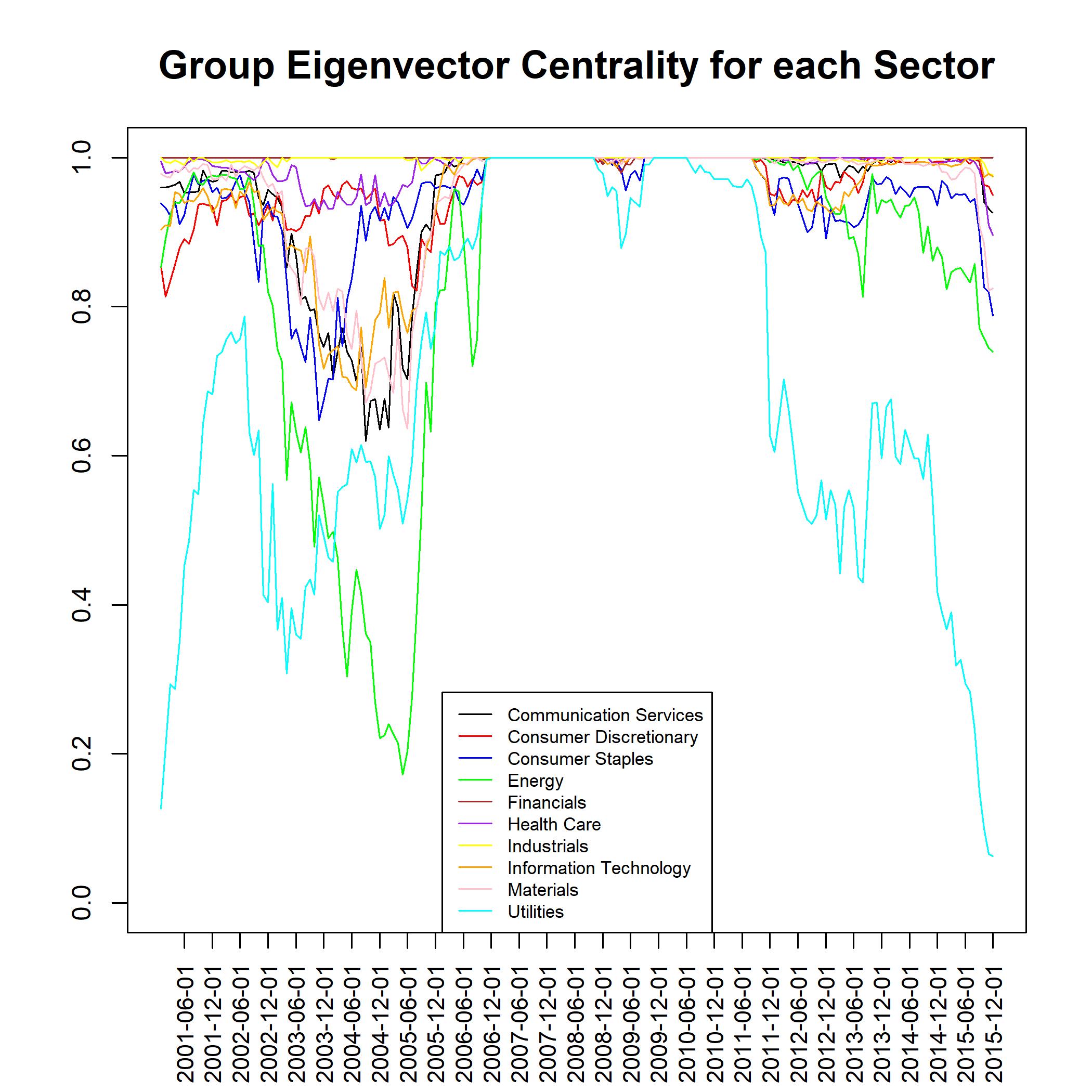}
	\includegraphics[scale=0.4]{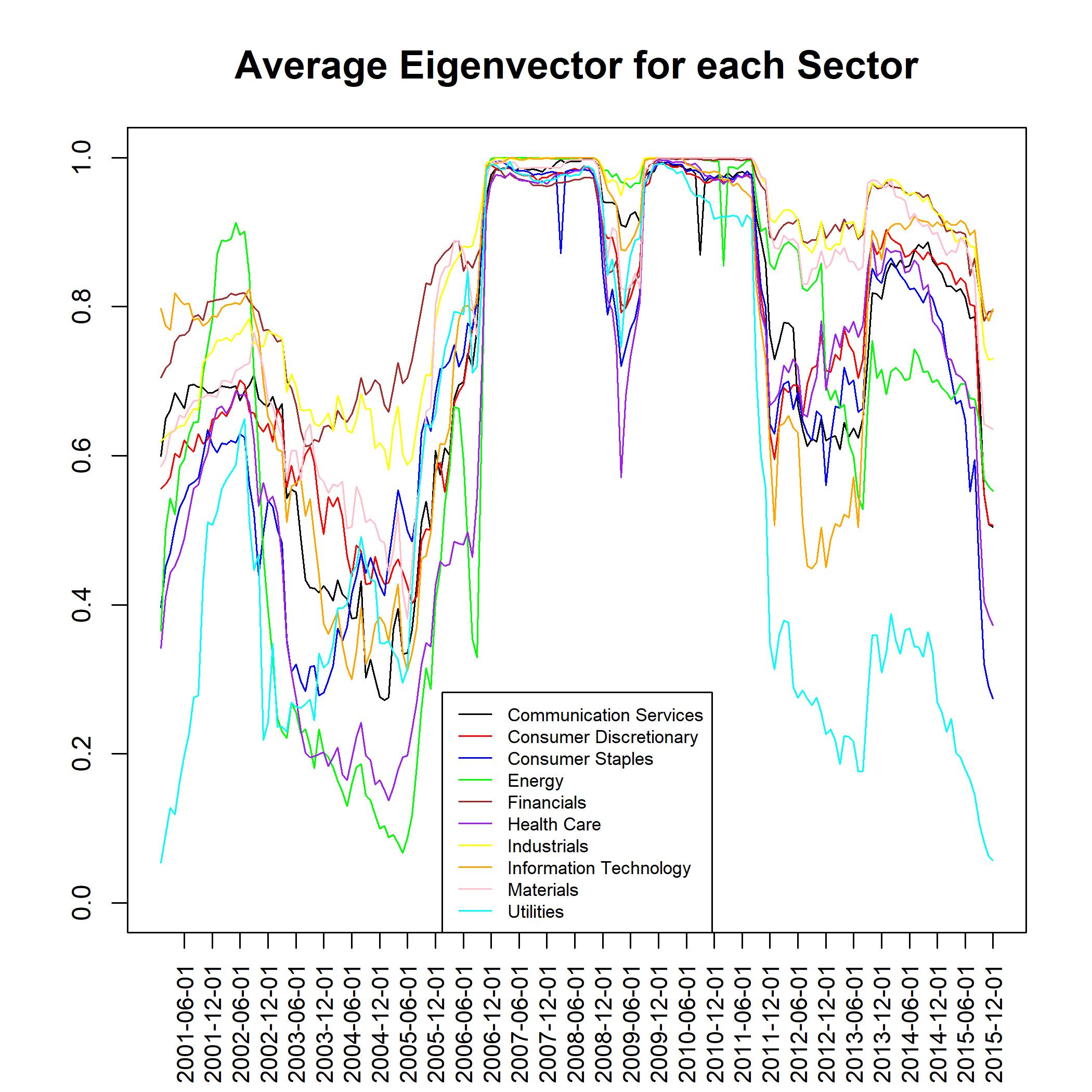}
	\caption{\added{We display, on the left side,} group eigenvector centrality $x(G_{s})$ \added{for each sector $s=1,...,10$ and for different time periods $t=1,...,181$. On the right side, we report the }average eigenvector centrality $\bar{x}(G_{s})$ for each sector $s=1,...,10$ and for different time periods $t=1,...,181$. }
	\label{fig:SecNetEigen}
\end{figure}

Moving to the eigenvector centrality, Figure \ref{fig:SecNetEigen} reports both the average eigenvector and the group eigenvector centrality, according to formula (\ref{releigen3}). Unlike the betweenness measure, the average centrality of sectors seems to be more reactive than the group centrality during the periods of crisis.\\
Since the eigenvector centrality quantifies the power of the group, it could be expected that this measure is particularly sensitive to a crisis of the system. Usually associated with a financial crisis, as for those of 2007-2008, contagion can be manifested as negative externalities diffused among the entities of the system.
Actually, the eigenvector group centrality is computed by means of the reduced model, where all the nodes of the same sector collapse in a single one leading to smooth differences between assets. Therefore, all sectors overreact moving towards the maximum value of centrality during the crisis periods, with the exception of the utilities' sector. A more remarkable behaviour is the one of the average centrality, where specific individual vertex centrality prevails on the sector, driving the trend. In particular, Financial, Industrial and Energy sectors seem to be more influenced than others, like Consumer Staples and Health Care.

Now we focus on the individual relative eigenvector centrality,  ranking the top and the bottom 15 (see Tables \ref{tab:4} and \ref{tab:5}, respectively), referred to the last period available (i.e. the network  \lq\lq1-2016\rq\rq). In particular, Table \ref{tab:4} reports the top 15 assets with highest relative eigenvector centrality in the period. The most part of the top ranked companies belongs to the financial sector. Financial companies appear as the most powerful one, both at individual level and as a part of a powerful group. Although such a sector has been significantly affected by the financial crisis, it has reinforced its position and dominates the international scene. Honeywell International is the only non-financial asset that belongs to the top 5 in terms of both betweenness and eigenvector centrality. This asset has been characterized during the period by a significant increase in the market value.\\
Table \ref{tab:5} lists, on the contrary, the 15 less central assets. The bottom of the ranking shows heterogeneity of the sectors: Utilities, Consumer Staples, Communication services and  Health Care are among the less central sectors. The main difference with respect to relative betweenness (Table \ref{tab:2}) is represented by the presence of several assets of Consumer Staples sector, probably reflecting a difficult period of the whole sector after the world crisis. In particular, the last two positions are occupied by Walmart and Target, that show a relative value significantly lower than all the other firms. At the end of 2015, Walmart saw its stock falling of 10\%\footnote{See for instance ``Wal-Mart Heirs See 11 Billion Vanish in a Day on Share Fall'', available at: https://web.archive.org/web/20151017210100/http://www.bloomberg.com/news/articles/2015-10-14/wal-mart-heirs-see-9-billion-vanish-in-a-day-as-shares-plummet}. In 2016, such a firm reported its first annual sales decline since 1980 and announced the closure of several stores. In 2017, Target Corporation shares suffered their biggest-ever price drop in active trade, as the discount retail giant struggles to cope with the “rapidly changing” behaviour of consumers. Target Corporation reported a profit that missed expectations and well below analyst projections. \\
Negative trends are also observed for other assets belonging to the 15 bottom ranked companies, for both eigenvector and betweenness centralities. For instance, the shares of the retail pharmacy giant CVS Health felt by more than 18\% during 2016, according to data from S\&P Global Market
Intelligence. Bristol-Myers stock collapsed in 2017 after a disastrous cancer study failure\footnote{Bristol-Myers has been the undisputed leader in immunotherapy, a new field of medicine that turns the body into a weapon against cancer. The company felt more than 16\% after the company announced that its drug, Opdivo, had failed to significantly boost the amount of lifetime and quality of life of a type of lung cancer patients, compared to chemotherapy.}.
A combination of bad news and a general sell-off in the stock market is sending the Allergan stock down since 2016. What is most surprising with the drop is the rapidity of its decline with the market value almost halved in two years. \\
It is worth noting the case of Exelon and other assets of Utility sector. These assets are more central in the subgroup than in the whole market. Indeed, Exelon -- which is one of the leader companies among the energy providers in the U.S. Utility sector -- is instead not very central in the network; at the same time, its centrality is not so affected by the crisis, probably due to a lower dependence between this sector and the other ones.

\begin{table}[!h]
	\begin{tabular}{|lc|c:ccc|}
		\hline\hline
		Asset Name & Sector & $x(i|G)$ & $x^{G}(i|G_s)$ & $k^{x}_{G_s}$ & $r^{x}_{G_s}$ \\
		\hline\hline
		
		Berkshire Hathaway &	Financials&	1.800	&1.000	&1.260	&1.428 \\
		Blackrock &	Financials	&1.751	&0.973&	1.260	&1.428 \\
		Honeywell International &	Industrials	&1,723	&0,980	&1,338	&1,314\\
		U.S. Bancorp	& Financials	&1,719	&0,955	&1,260	&1,428\\
		JP Morgan &	Financials	&1,695	&0,942	&1,260	&1,428 \\
		Citigroup &	Financials	&1,691	&0,940	&1,260	&1,428\\
		Visa &	Financials &	1,687	&0,937	&1,260	&1,428\\
		Mastercard &	Financials	&1,687	&0,937	&1,260	&1,428\\
		Bank of New York Mellon  &	Financials	&1,684	&0,936 &1,260	&1,428\\
		Morgan Stanley &	Financials	&1,684	&0,936	&1,260	&1,428\\
		Intel  &	Information Technology&	1,662	&0,948	&1,223	&1,434\\
		Goldman Sachs &	Financials&	1,659&	0,922	&1,260	&1,428\\
		Bank of America &	Financials&	1,650&	0,917&	1,260&	1,428\\
		Fedex &	Industrials	&1,640&	0,932&	1,338&	1,314\\
		Texas Instruments &	Information Technology&	1,623	&0,925&	1,223&	1,434\\
				\hline\hline
	\end{tabular}
	\caption{Top 15 ranking in terms of relative eigenvector based on the network \lq\lq 1-2016\rq\rq}
	\label{tab:4}
\end{table}

\begin{table}[!h]
	\begin{tabular}{|lc|c:ccc|}
		\hline\hline
		Asset Name & Sector & $x(i|G)$ & $x^{G}(i|G_s)$ & $k^{x}_{G_s}$ & $r^{x}_{G_s}$ \\
		\hline\hline
		
Charter Communications	&Communication Services	& 0.247	& 0.148 &	1.835	&0.908 \\
Verizon Communications &	Communication Services&	0.242 &0.134 &	1.260 &	1.428 \\
Eli Lilly and Company	&Health Care	&0.181 &	0.112 &	2.403 &	0.671 \\
Simon Property Group	&Financials	&0.178 &	0.099 &	1.260 &	1.428\\
Altria Group	&Consumer Staples	&0.126 &	0.089 &	2.872 &	0.494\\
Exelon 	&Utilities	&0.119	&1.059 &	1.087 &	0.104\\
Allergan Plc	&Health Care	&0.112	&0.070	&2.403	&0.671\\
Nextera Energy	&Utilities	&0.106 &0.941 &	1.087 &	0.104\\
Duke Energy 	&Utilities	&0.097	&0.863	&1.087	&0.104\\
Costco Wholesale	&Consumer Staples	&0.093	&0.066 &	2.872	&0.494\\
Southern Company	&Utilities	&0.092	&0.818 &	1.087 &	0.104\\
Bristol-Myers Squibb Company	&Health Care	&0.089	&0.055	&2.403	&0.671\\
CVS Health 	&Health Care	&0.088 &	0.054 &	2.403 &	0.671\\
Target &Consumer Staples	&0.026 &	0.018 &	2.872 &	0.494\\
Walmart	&Consumer Staples	&0.014	&0.010	&2.872	&0.494\\
				\hline\hline
	\end{tabular}
\caption{Bottom 15 ranking in terms of relative eigenvector based on the network \lq\lq 1-2016\rq\rq}
\label{tab:5}
\end{table}

\section{Conclusions and future research}

The paper contains a new conceptualization of centrality measures which includes also the role of the single nodes and of the subgraphs of the network in the overall system. The scientific ground of the study lies in the need of exploring the relative relevance of
such elements of a complex network in their real contextualization. The definition of relative centrality measures allows to compare nodes and subgraphs, also when they belong to different networks.

After a theoretical description of the model, some empirical experiments have been carried out. The employed dataset consists of the components of the $S\&P$ 100 index, which are assumed to be connected through their correlation coefficients. \deleted{Two relative centrality measures have been tested: eigenvector and betweenness centrality.}
\added{We focused on two measures (eigenvector and betweenness centrality) with extremely different characteristics, in order to discover the hidden role of influential firms in local groups. Results show that both measures provide additional insight than the simple degree centrality, that assures only a local view}. \color{black}
From a general point of view, we detected an homogeneous behaviour of the relative centralities in all sectors during the period of crisis, in response to the increase of the assets correlation. The financial sector, that has suffered most, due to the effects of the crisis, has returned to have a powerful role and it prevails in conveying information. Industrial and Energy sectors also have increased their importance in terms of power and  dominance. Centralities of assets and sectors not always go accordingly. Assets in some sectors, as Consumer Staple, Utilities and Communication, have diminished their importance, probably also reflecting how specific firms evolved over time.
Results highlight the centrality of specific stocks (nodes) or sectors (subgraphs) in the overall system, and relevant insights have been derived under a purely economic point of view. \\
\added{However, it is noteworthy that our proposal of decomposition of a relative centrality can be easily extended also to other relevant measures,} 
\noindent \added{as well as to oriented networks, in order to assess the emergence of opinion leaders at different levels for example in voting models. 
Moreover, the introduced methodological tools can be effectively applied to other relevant empirical data. An important example is the world trade network. In this specific case, relative and group centralities might be of interest for detecting the economic trading flows within the overall world context.}
\added{In light of the possible applications of the presented centrality measures, we also point out the crucial role of such devices in describing the time-evolving properties of the networks topology. In this respect, we here deal with static measures on rolling time-windows, which can give insights on time-evolution when computed over different time periods. However, an extension to a dynamic setting can be of interest. To this aim, one should understand the dynamics underlying the evolution of relative and groups centrality measures by assessing the presence of regularities in the relationship between different time-realizations. Such an evolutionary rule would be reasonably of random nature, able to describe the future evolution of networks topology. This topic might contribute to effectively predict crucial economic and financial patterns, like the world trade and the financial stock markets.}

%\added{In the empirical application, we focused on two measures with extremely different characteristics, in order to discover the hidden role of influential companies in local groups. Both measures are more informative than the simple degree centrality, that is characterized as a local measure. However, formulas \ref{rel2} and \ref{rel3} can be also applied to other measures.}

\section*{Compliance with Ethical Standards:}

%(In case of Funding) Funding: This study was funded by X (grant number X). 

Conflict of Interest: The authors declare that there is no conflict of interest.\\
%(In case animals were involved) Ethical approval : All applicable international, national, and/or institutional guidelines for the care and use of animals were followed.
%(And/or in case humans were involved) Ethical approval: All procedures performed in studies involving human participants were in accordance with the ethical standards of the institutional and/or national research committee and with the 1964 Helsinki declaration and its later amendments or comparable ethical standards.
Ethical approval: This article does not contain any studies with human participants or animals performed by any of the authors.
%(Or) Ethical approval: This article does not contain any studies with animals performed by any of the authors.
%(Or) Ethical approval: This article does not contain any studies with human participants or animals performed by any of the authors.

%(In case humans are involved) Informed consent:  Informed consent was obtained from all individual participants included in the study.

% BibTeX users please use one of
\bibliographystyle{spbasic}      % basic style, author-year citations
%\bibliographystyle{spmpsci}      % mathematics and physical sciences
%\bibliographystyle{spphys}       % APS-like style for physics
%\bibliography{Myref}   % name your BibTeX data base

% Non-BibTeX users please use
%\begin{thebibliography}{}

%
% and use \bibitem to create references. Consult the Instructions
% for authors for reference list style.
%
%\bibitem{RefJ}
% Format for Journal Reference
%Author, Article title, Journal, Volume, page numbers (year)
% Format for books
%\bibitem{RefB}
%Author, Book title, page numbers. Publisher, place (year)
% etc
%\end{thebibliography}

%\section{Appendix}

\end{document}